\documentclass[prd,nofootinbib,floatfix,preprintnumbers, twocolumn, superscriptaddress]{revtex4}%
\pdfoutput=1 

\usepackage{amssymb,amsmath}
\usepackage{color}
\usepackage{graphicx}
\usepackage{comment}
\usepackage{epsfig}
\usepackage{latexsym} % Provided \Box\begin{figure}[htbp]
\usepackage{multirow}

\def\simge{%  ``greater than about'' symbol
    \mathrel{\rlap{\raise 0.511ex
        \hbox{$>$}}{\lower 0.511ex \hbox{$\sim$}}}}
\def\simle{%  ``less than about'' symbol
    \mathrel{\rlap{\raise 0.511ex
        \hbox{$<$}}{\lower 0.511ex \hbox{$\sim$}}}}
\def\beq{\begin{equation}}
\def\eeq{\end{equation}}
\def\barr{\begin{eqnarray}}
\def\earr{\end{eqnarray}}
\def\bc{\begin{center}}
\def\ec{\end{center}}
\def\tw{\textwidth}
\newcommand{\ig}{\includegraphics}

\begin{document}

\title{The Flavor Structure of the Excited Baryon Spectra from Lattice QCD}

\author{Robert~G.~Edwards}
\email{edwards@jlab.org}
\affiliation{Jefferson Laboratory, 12000 Jefferson Avenue,  Newport News, VA 23606, USA}

\author{Nilmani~Mathur}
\email{nilmani@theory.tifr.res.in}
\affiliation{Department of Theoretical Physics, Tata Institute of Fundamental Research, Homi Bhabha Road, Mumbai 400005, India}

\author{David G. Richards}
\email{dgr@jlab.org}
\affiliation{Jefferson Laboratory, 12000 Jefferson Avenue,  Newport News, VA 23606, USA}

\author{Stephen J.~Wallace}
\email{stevewal@umd.edu}
\affiliation{Department of Physics, University of Maryland, College Park, MD 20742, USA}

\collaboration{for the Hadron Spectrum Collaboration}
\date{December 20, 2012}
%\date{\today}

\begin{abstract}
 Excited state spectra are calculated using lattice QCD for baryons that 
can be formed from $u$, $d$ and $s$ quarks, namely the $N$, $\Delta$, $\Lambda$, 
$\Sigma$, $\Xi$ and $\Omega$ families of baryons.   
Baryonic operators are constructed from continuum operators that transform as 
irreducible representations of SU(3)$_F$ symmetry for flavor, SU(4) 
symmetry for Dirac spins of quarks and O(3) symmetry for orbital angular 
momenta.  Covariant derivatives are used to realize orbital angular 
momenta. Using the operators, we calculate matrices of correlation functions
in order to extract excited states. The resulting lattice spectra 
have bands of baryonic states with well-defined total spins up to $J=\frac{7}{2}$. 
Each state can be assigned a dominant flavor symmetry and the counting of states of each flavor and spin reflects 
$SU(6) \times O(3)$ symmetry for the lowest negative-parity and positive-parity bands.  
States with strong hybrid content are identified through the dominance of chromo-magnetic operators.
\end{abstract}

\maketitle

\section{Introduction}\label{sec:intro}%
The spectra of baryon resonances have been a focus of study
experimentally and theoretically, both for particles composed of the
light $u$ and $d$ quarks, and those containing one or more of the heavy $c$
and $b$ quarks.  There is increasing activity aimed at
understanding the spectra of particles containing one or more $s$
quarks, so-called hyperon physics, e.g., \cite{Aubert:2006dc}.  
At present the knowledge of the $\Xi$ and $\Omega$ families 
is particularly limited with only a few states
experimentally established, and scant knowledge as to their
properties~\cite{Beringer-PDG:2012}. 

A quantitative description of the spectra of baryons that can
be constructed from the $u$, $d$ and $s$ quarks is important for
a number of reasons.  Firstly, to explore how the effective degrees of
freedom that describe the hadron spectra change as the masses of the
quarks are changed.  Secondly, to advance our understanding of QCD in
regimes where the hyperons play a crucial role, such as in the physics
of the early universe and core-collapse in supernovae. Finally, to address 
the long-term goal of extracting baryon spectra from lattice QCD.

In this paper, we present lattice QCD calculations of the
excited-state spectra of baryons that can be constructed from
$u$, $d$ and $s$ quarks: the $N$, $\Delta$, $\Lambda$, $\Sigma$, $\Xi$
and $\Omega$ families of baryons.  Our calculations exploit a menu of methods that we
have developed and are key to our studies of excited-state spectra: the
use of an anisotropic, clover action for the generation of gauge configurations~\cite{Edwards:2008ja,Lin:2008pr}, the construction of
operators that respect the symmetries of the lattice yet which retain
a memory of their continuum analogues~\cite{Edwards:2011jj}, the 
use of ``distillation''~\cite{Peardon:2009gh} to efficiently 
compute the correlation functions between those operators,
and finally the application of the variational 
method~\cite{Michael:1985,LuscherWolff:1990}, exploiting the
eigenvectors of matrices of correlation functions to determine the spins and flavors of the extracted
energies.  In two earlier works,~Refs.~\cite{Edwards:2011jj,Dudek:2012ag}, we applied these methods to compute
the nucleon and $\Delta$ spectra, with the spins of the states 
identified clearly; these works revealed spectra at least as rich as the
quark model, with suggestions of ``hybrid'' states in which the gluons
played an important structural role.  Here we expand upon these
works by considering baryons containing not only the light $u$ and $d$ quarks, but
also one or more $s$ quarks.

The layout of the remainder of the paper is as follows.  In the next
section, we describe details of the lattices used, and outline the
construction of the lattice interpolating operators.  
In Sec.~\ref{sec:fits}, we recall our procedure for analyzing the
hadron correlation functions, and for identifying the continuum spins of
the states. Our
results are presented in Sec.~\ref{sec:results} and a summary of the work is
presented in Sec.~\ref{sec:summary}.

\section{Computational Methods and Details \label{sec:details}}
A significant challenge in determining the excited-state spectra is
to obtain a sufficient number of energy levels in order to extract their 
patterns of energies, spins and flavors
before statistical noise overcomes the signal.  This requires
accurately resolving the behavior of hadron correlation functions at
short temporal separations.  A computationally efficient calculation
is obtained through the use of an anisotropic action, with a finer
temporal lattice spacing than that used in the spatial directions,
enabling correlation functions to be resolved over several time slices
while preserving a sufficient spatial volume.  The lattice action used in
this work, as well as the method used to tune the parameters of the
action, can be found in Refs.~\cite{Edwards:2008ja,Lin:2008pr}.  To summarize, we use improved
gauge and fermion actions with two mass-degenerate light quarks of
mass $m_l$ and a strange quark of mass $m_s$.  We employ $16^3 \times
128$ lattices having spatial lattice spacing $a_s \sim 0.123~{\rm fm}$,
and a renormalized anisotropy, the ratio of the spatial and temporal
lattice spacings, of $\xi \approx 3.5$.  The calculation is performed at three
values of the light-quark masses, corresponding to pion masses of $391$,
$524$ and $702~{\rm MeV}$. The $702~{\rm MeV}$ pion mass
corresponds to the $SU(3)_F$ 
flavor-symmetric point.
Some details of the
lattices are summarized in Table~\ref{tab:lattices}.  The mass of the $\Omega$ is used
to set the scale. It was determined within an estimated uncertainty of 2\% in Ref.~\cite{Bulava:2010yg}
on the same ensembles.
\begin{table}[t]
  \begin{tabular}{cccccccc}
    $a_t m_\ell$ & $a_t m_s$ & $m_\pi$ & $m_K/m_\pi$ & \multicolumn{1}{c|}{$a_t m_\Omega$} & $N_{\mathrm{cfgs}}$ & $N_{\mathrm{t_{srcs}}}$ & $N_{\mathrm{vecs}}$ \\
    \hline
     $-0.0743$ & $-0.0743$ & 702 & 1.00  & $0.3593(7)$ &  500 & 7 & 56 \\
     $-0.0808$ & $-0.0743$ & 524 & 1.15 & $0.3200(7)$ &  500 & 7 & 56 \\
$-0.0840$ & $-0.0743$ & 391 & 1.39 & $0.2951(22)$ & 479 & 8 & 56 \\
  \end{tabular}
  \caption{Parameters of the $16^3 \times 128$ lattices and propagators used in this work.
    The pion mass in MeV, and the number of configurations are listed, as well as the number of time-sources and the number of distillation vectors $N_{\mathrm{vecs}}$.}
\label{tab:lattices}
\end{table}

\subsection{Baryon Interpolating operators}
The construction of the baryon interpolating operators is described in
Ref.~\cite{Edwards:2011jj}.  Our starting point is to construct a
set of continuum baryon interpolating operators, which we express symbolically as,
\begin{equation}
{\cal O}^{J^P} \sim  \left( F_{\Sigma_F} \otimes (S^{P_s})_{\Sigma_S}^n \otimes D^{[d]}_{L,
      \Sigma_D} \right)^{J^P},
\end{equation}
where the factors $F, S~{\rm and}~D$ describe the flavor, Dirac spin
and derivative structure of the operators, respectively, and the
subscripts, $\Sigma_i$, denote the corresponding symmetry
representations with respect to permutations.
The Dirac spin
factor, $(S^{P_S})^n_{\Sigma_S}$, represents the combination of three
Dirac-spinor quark fields with overall angular momentum $S$ and parity $P_S$, with
permutational symmetry $\Sigma_S$, while $n$ labels different
constructions.  The spatial factor, $D^{[d]}_{L, \Sigma_D}$, is
expressed in terms of gauge-covariant derivatives acting on the three
quark fields, where $[d]$ denotes the number of derivatives, $\Sigma_D$ is
the permutation symmetry, and the derivatives are combined so as to transform as
angular momentum $L$.  Finally, the factors are combined to yield an
interpolator of definite spin and parity, which we label $J^P$.

The construction of operators is guided by symmetry considerations. Based upon
four-component Dirac-spinor quark fields combined with $SU(3)$ flavor plus angular momentum, 
the symmetry of the full set
of operators is $SU(12)\times O(3)$. An important subset of operators is based on
non-relativistic quark spins, by which we mean operators based on the upper two components of Dirac-spinor quark fields.
In this study, we realize angular momenta by including up to two covariant derivatives, $d = 0,
1, 2$, with maximum accessible values of orbital angular momentum, $L$, of $0, 1,\, {\rm and} \,
2$, respectively.  Tables~\ref{tab:D1_SU6_O3} and \ref{tab:D2_SU6_O3}
show the patterns of quantum numbers of operators based on non-relativistic spins. 
Not listed are
operator constructions based on the lower components of Dirac spinors 
(relativistic quark spins), although they outnumber the non-relativistic operators
 in the set used. 

Three non-relativistic quark spins are mixed-symmetric 
for $S=\frac{1}{2}$ and symmetric for $S=\frac{3}{2}$. Orbital angular momenta 
are mixed-symmetric for $L=1$ and symmetric, mixed-symmetric or antisymmetric 
for $L=2$.  The combination of flavor, spin, and space symmetries must be 
symmetric to yield an operator that is antisymmetric when color is included,
in accord with the Pauli Principle.  
We list in the tables the spin-parity, $J^P$, of the 
distinct combinations that are allowed. 
The classification of operators according to $SU(3)_F$ flavor symmetry 
will be exploited as a means of identifying the dominant flavor structure
of the states in the extracted spectrum.

The operators constructed from non-relativistic quark spinors
have the spin, parity and flavor quantum numbers allowed by $SU(6)\times O(3)$ symmetry.
 These quantum numbers occur
in definite patterns that are indicated by the bold numbers
in Tables~\ref{tab:D1_SU6_O3} and \ref{tab:D2_SU6_O3}, i.e.,  
there are $N_1(J)$, 
$N_8(J)$ and $N_{10}(J)$ operators for SU(3)$_F$ singlet, octet and
decuplet symmetries, respectively, with total angular momentum $J$.
If the states in the spectra were to correspond to broken 
$SU(6)\times O(3)$ symmetry,
then there would be patterns with the same numbers of 
states and the same quantum numbers. 
We will compare our spectra with the 
$SU(6)\times O(3)$ patterns given in Tables~\ref{tab:D1_SU6_O3} and \ref{tab:D2_SU6_O3}.

For $d=2$, the mixed-symmetry combination
$D^{[2]}_{L=1,M}$ corresponds to the commutator of two covariant
derivatives, producing a chromo-magnetic field,
as explained in Ref.~\cite{Dudek:2012ag}.  
We call these hybrid operators because of their essential gluonic
content.  
The hybrid operators based on non-relativistic quark spinors are listed 
in Table~\ref{tab:D2_hybrid}. The patterns of states expected to be created by such hybrid operators
correspond to the numbers $M_1(J)$, $M_8(J)$ and $M_{10}(J)$.

\begin{table}
\begin{tabular}{|ccc|cccc|}
\hline
$SU(3)_F$ & ~~S~~ & ~~L~~ &  & ~$J^{P}$~ & ~~ ~~ & ~~ ~~\\
\hline
  & & & & & &  \\
${\bf 8_F}$ & $\frac{1}{2}$ & 1 &  $ \frac{1}{2}^-$ & $\frac{3}{2}^-$ & &   \\
 &  $\frac{3}{2}$ & 1 & $ \frac{1}{2}^-$ & $ \frac{3}{2}^-$ & $ \frac{5}{2}^-$ &  \\
  & & & & & &  \\
\hline
$N_8(J)$ & & & {\bf 2} & {\bf 2} & {\bf 1} & \\
 \hline
  & & & & & &    \\
     ${\bf 10_F}$ & $\frac{1}{2}$ & 1  &  $ \frac{1}{2}^-$ &$\frac{3}{2}^-$ & &  \\
 & & & & & &  \\
 \hline
$N_{10}(J)$ & & &  {\bf 1} & {\bf 1} & {\bf 0} & \\
    \hline
  & & & & & &  \\ 
       ${\bf 1_F}$ &  $\frac{1}{2}$ & 1  &  $ \frac{1}{2}^-$ & $\frac{3}{2}^-$ & & \\
 & & & & & &  \\
 \hline
$N_1(J)$ & & &  {\bf 1} & {\bf 1} & {\bf 0} & \\
 \hline     
 \end{tabular}
\caption{Allowed spin-parity patterns for one-derivative operators based on
 non-relativistic quark spinors ($SU(6) \times O(3)$ symmetry).
For each flavor, the quark spin S and orbital angular momentum L are listed
followed by the allowed $J^P$ values.
The total number of operators is listed
as $N_8$ for flavor octets, $N_{10}$ for flavor decuplets and $N_1$ for flavor singlets.
\label{tab:D1_SU6_O3} } 
\vspace{0.1in}
 \end{table}

 \begin{table}[h]
\begin{tabular}{|ccc|cccc|}
\hline 

$SU(3)_F$ & ~~S~~ & ~~L~~ & ~~ ~~  & ~$J^{P}~ $ ~~ ~~  & ~~ ~~ &  \\ 
\hline 
& & & & & &  \\
${\bf 8_F}$ & $\frac{1}{2}$ & 0 &  $ \frac{1}{2}^+$ & & &    \\
 & $\frac{1}{2}$ & 0 &   $ \frac{1}{2}^+$ & & &    \\
 &  $\frac{1}{2}$ & 1 &  $ \frac{1}{2}^+$ & $ \frac{3}{2}^+$& &   \\
 &  $\frac{1}{2}$ & 2 &  & $ \frac{3}{2}^+$ & $ \frac{5}{2}^+$ &   \\
 &  $\frac{1}{2}$ & 2 &  & $ \frac{3}{2}^+$ & $ \frac{5}{2}^+$ &   \\
 &   $\frac{3}{2}$ & 0 & & $ \frac{3}{2}^+$ & &  \\ 
 &   $\frac{3}{2}$ & 2 &  $\frac{1}{2}^+$ & $\frac{3}{2}^+$ & $\frac{5}{2}^+$ & $\frac{7}{2}^+$ \\ 
& & & & & &  \\
 \hline
$N_8(J)$ & & & {\bf 4} & {\bf 5} & {\bf 3} & {\bf 1} \\
    \hline
    & & & & & &  \\
    ${\bf 10_F}$ &  $\frac{1}{2}$ & 0 &  $ \frac{1}{2}^+$ & & &  \\
&  $\frac{1}{2}$ & 2 &   & $ \frac{3}{2}^+$ & $ \frac{5}{2}^+$&     \\
 & $\frac{3}{2}$ & 0 &  &  $ \frac{3}{2}^+$ & &  \\
    &  $\frac{3}{2}$ & 2 & $\frac{1}{2}^+$ & $\frac{3}{2}^+$ & $\frac{5}{2}^+$ & $\frac{7}{2}^+$ \\
     & & & & & &  \\
 \hline
 $N_{10}(J)$    & & & {\bf 2} & {\bf 3} & {\bf 2} & {\bf 1}  \\
    \hline
   & & & & & &  \\
${\bf 1_F}$ & $\frac{1}{2}$ & 0 & $ \frac{1}{2}^+$ & & & \\
   & $\frac{1}{2}$ & 2 &  & $ \frac{3}{2}^+$ & $ \frac{5}{2}^+$&   \\
   &  $\frac{3}{2}$ & 1 &   $\frac{1}{2}^+$ & $\frac{3}{2}^+$ & $\frac{5}{2}^+$ & \\
   & & & & & &  \\
 \hline
  $N_1(J)$   & & & {\bf 2} & {\bf 2} & {\bf 2} & {\bf 0}  \\
    \hline
\end{tabular}
\caption{Allowed spin-parity patterns for two-derivative operators 
based on non-relativistic quark spinors ($SU(6)\otimes O(3)$ symmetry). 
For ${\bf 8_F}$, two distinct operators with different internal symmetries are allowed 
for L=0 combined with S=$\frac{1}{2}$ and for L=2 combined with S=$\frac{1}{2}$. 
Hybrid 
operators are listed separately in Table~\ref{tab:D2_hybrid}.
The total number of operators for each spin is listed
as $N_8$ for flavor octets, $N_{10}$ for flavor decuplets and $N_1$ for flavor singlets.
\label{tab:D2_SU6_O3} } 
\end{table}

 \begin{table}[h]
\begin{tabular}{|ccc|ccc|}
\hline 

$SU(3)_F$ & ~~S~~ & ~~L~~ &  & ~$J^{P}$~ & ~~ ~~   \\ 
\hline 
& & & & &  \\
${\bf 8_F}$ & $\frac{1}{2}$ & 1 &  $ \frac{1}{2}^+$ & $\frac{3}{2}^+$ &     \\
 & $\frac{3}{2}$ & 1 &  $ \frac{1}{2}^+$ & $\frac{3}{2}^+$ & $\frac{5}{2}^+$   \\
 &&&&& \\
 \hline
$M_8(J)$ & & & {\bf 2} & {\bf 2} & {\bf 1}  \\
    \hline
    & & & & &  \\
    ${\bf 10_F}$ & $\frac{1}{2}$ & 1 & $\frac{1}{2}^+$  &$\frac{3}{2}^+$ &   \\
 &&&&&\\
 \hline
 $M_{10}(J)$    &  &  & {\bf 1} & {\bf 1} & {\bf 0}  \\
    \hline
    & & & & &  \\
    ${\bf 1_F}$ & $\frac{1}{2}$ & 1 & $\frac{1}{2}^+$  &$\frac{3}{2}^+$ &   \\
  &&&&&\\
 \hline
  $M_1(J)$    & & & {\bf 1} & {\bf 1} & {\bf 0}    \\
    \hline
\end{tabular}
\caption{Allowed spin-parity patterns for hybrid two-derivative operators based on non-relativistic 
quark spinors, following Ref.~\cite{Dudek:2012ag}.  The operators correspond to 
the combination of three quarks in a color octet with a gluonic field, $G$, to make a 
color singlet as $\big[ (qqq)_{{\bf 8}_c} G_{{\bf 8}_c}\big]_{{\bf 1}_c}$. The 
gluon field has spin-parity $1^+$.
The patterns of states expected to be created through the coupling of a 
chromo-magnetic gluon field coupled to quark fields in a color octet 
correspond to the numbers $M_8$ for flavor 
octets, $M_{10}$ for flavor decuplets and $M_1$ for flavor singlets.  The 
counting of operators is the same as in Table~\ref{tab:D1_SU6_O3} but the operators
have reversed parity.
\label{tab:D2_hybrid} } 
\end{table}

The final step in the construction of the interpolating operators is
the subduction to the irreducible representations (irreps) of the cubic group.
See Ref.~\cite{Edwards:2011jj} for details.
The number of operators used in each lattice irrep
is the same for both positive and negative parities, which are
denoted by the subscripts $g$ and $u$, respectively, and the operators 
are classified according to the flavor
irreps of $SU(3)_F$, as shown in Table~\ref{tab:su2flavor}.  

In order to reduce the complexity of the analysis, some relativistic 
operators are omitted.  We include all non-relativistic
operators based on the upper-component Dirac spinors.
For example, for $\Sigma$, 68 operators 
in irrep $H$ are used, with 37 having flavor octet symmetry and 31 
having flavor decuplet symmetry. This set omits 22 relativistic operators 
but is sufficient to determine the spectrum. 
Because we omit some relativistic operators, and we include 
the ``hybrid'' operators introduced in Ref.~\cite{Dudek:2012ag}, the 
number of operators differs from Ref.~\cite{Edwards:2011jj}.

The spectra for the baryons made from $u$ and $d$ quarks, i.e., $N$ and $\Delta$, 
have been explored in earlier works using the same
ensembles and operator constructions we will employ here, firstly in a calculation of the spectra
for both parities but with the ``hybrid'' operators excluded from the
basis\cite{Edwards:2011jj}, and then of the positive-parity spectra
using all operators, including the  ``hybrid'' operators\cite{Dudek:2012ag}.  In this
work we consider excited states of both parities for baryons that
can be formed from $u$, $d$ and $s$ quarks.

\begin{table}
\begin{tabular}{c|c|cc|ccc}
  & $SU(3)_F$ & $I$ & $S$ & $G_1$ & $H$ & $G_2$  \\
\hline
$N$        & ${\bf 8_F}$  & $\frac{1}{2}$ & 0  & 22 & 37 & 15 \\
$\Delta$   & ${\bf 10_F}$ & $\frac{3}{2}$ & 0  & 19 & 31 & 12 \\
$\Lambda$  & ${\bf 1_F}$  & $0$           & 0  & 17 & 27 & 10 \\
$\Lambda$  & ${\bf 8_F}$  & $0$           & 0  & 22 & 37 & 15 \\
$\Sigma$   & ${\bf 8_F}$  & $1$           & -1 & 22 & 37 & 15 \\
$\Sigma$   & ${\bf 10_F}$ & $1$           & -1 & 19 & 31 & 12 \\
$\Xi$      & ${\bf 8_F}$  & $\frac{1}{2}$ & -2 & 22 & 37 & 15 \\
$\Xi$      & ${\bf 10_F}$ & $\frac{1}{2}$ & -2 & 19 & 31 & 12 \\
$\Omega$   & ${\bf 10_F}$ & $0$           & -3 & 19 & 31 & 12 \\
\hline
\end{tabular}
\caption{For each $SU(3)_F$ symmetry, isospin $I$ and strangeness $S$,
the numbers of operators are shown in the last three columns for each lattice irrep
  $G_{1}$, $H$ and $G_{2}$. The constructions use
  operators with up to two derivatives. The same numbers of operators
are used for both positive and negative parities.
\label{tab:su2flavor}}
\end{table}

\subsection{Correlator Analysis}\label{sec:fits}
The variational method we use in our analysis involves the computation
of a matrix of correlation functions,
\begin{equation}
C^{\Lambda}_{ij}(t \equiv t_f - t_i) = \langle 0 \mid {\cal O}_i(t_f) {\cal O}^{\dagger}_j(t_i) \mid
0 \rangle ,
\end{equation}
for operators $i$ and $j$ that lie in a given cubic irrep,
$\Lambda$. 
With the large operator basis
introduced above, it is essential to have an efficient computational
method for computing the baryon correlation functions.  We use
{\em distillation}~\cite{Peardon:2009gh}, which provides a smearing
function whilst enabling the full matrix of correlators to be
constructed for all the operators at both source and sink.  Following
our earlier work using these lattices, we employ $N=56$ eigenvectors
of the gauge-covariant Laplacian when constructing the distillation operator,
and compute correlation functions from $N_{\mathrm{t_{srcs}}}$ time sources, as
listed in Table~\ref{tab:lattices}.  

\begin{figure*}
     \ig[width=1.0\tw,height=1.0\tw]{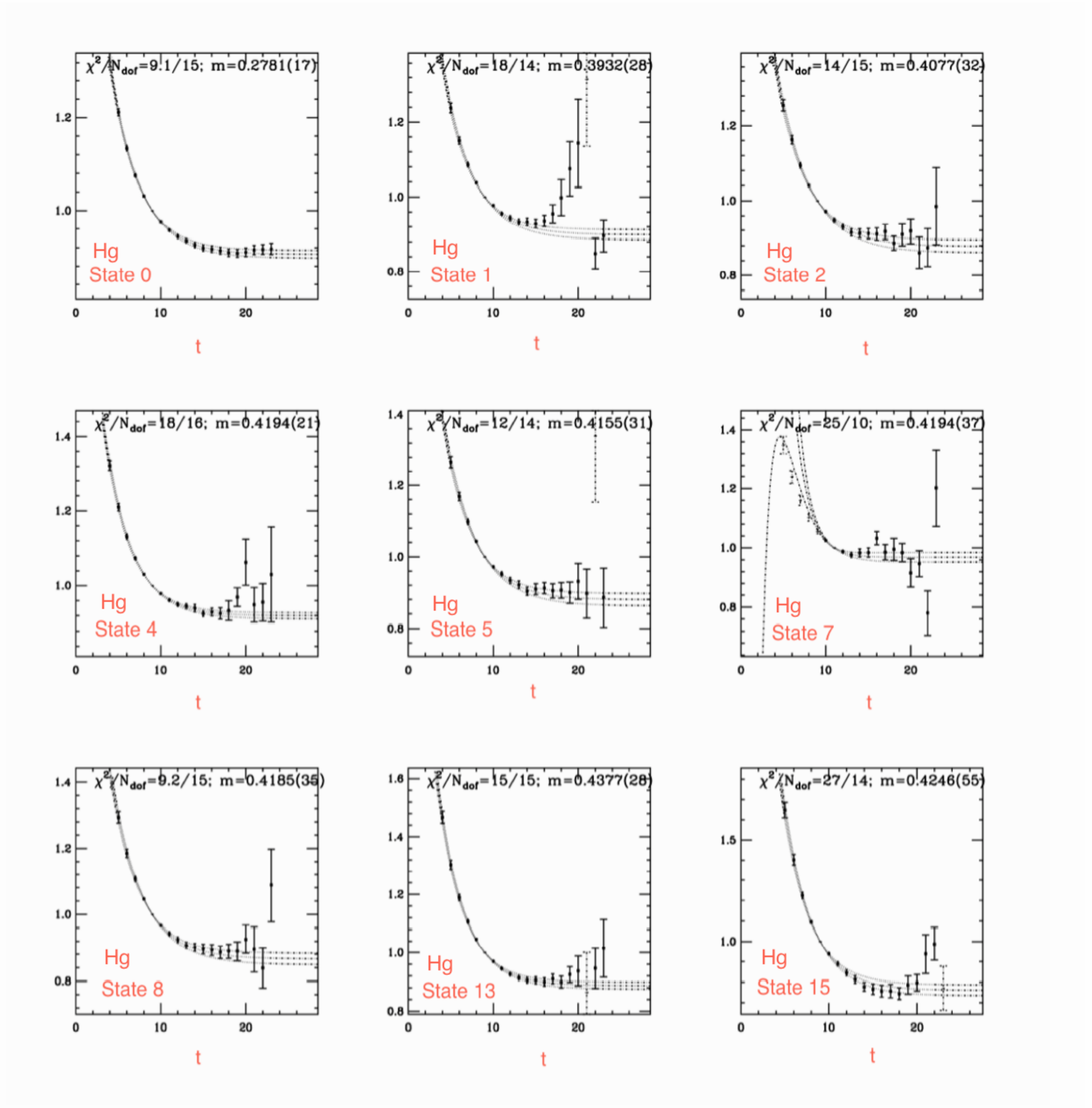} 
\caption{Fits to principal correlators for nine states in irrep $H_g$  that are identified as
 $J=\frac{3}{2}^+$. 
Fits are obtained using Eq.~(\ref{eq:fit}).  For 
plotting, we divide out the first exponential factor, thus, the plots show 
values of $e^{m_{\mathfrak n} (t-t_0)} \lambda_{\mathfrak n}(t)$
as the data points. Lines show the fits according to
the form $e^{m_{\mathfrak n} (t-t_0)} \lambda_\mathfrak{n}(t) =
1-A_\mathfrak{n} + A_\mathfrak{n}\,
e^{-(m'_{\mathfrak n}- m_{\mathfrak n}) (t-t_0)}$, with $t_0=9$.
 The fits approach the constant value, $1-A_\mathfrak{n}$, for large $t$.
The bands indicate the fits with the central fit values along with 
one-standard-deviation uncertainties.
	\label{fig:Sigma_840_PrinCorrPlots} }
	\end{figure*}

Our fitting strategy follows exactly that outlined in
Refs.~\cite{Edwards:2011jj} and \cite{Dudek:2012ag}.  In summary, we
solve the generalized eigenvalue equation,
\begin{equation}
C(t) v_{\mathfrak n}(t,t_0) = \lambda_{\mathfrak n}(t,t_0) C(t_0) v_{\mathfrak n}(t,t_0) \label{eq:gev} ,
\end{equation}
and thereby obtain the masses from the principal correlators,
$\lambda_{\mathfrak n}(t,t_0),\  {\mathfrak n} = 1\dots {\rm dim}(C)$, 
by fitting to,
\begin{equation}
\lambda_{\mathfrak n}(t,t_0) = (1 - A_{\mathfrak n}) e^{-m_{\mathfrak n}(t - t_0)} + A_{\mathfrak n} e^{-m'_{\mathfrak n}(t - t_0)}.
\label{eq:fit}
\end{equation}
The corresponding eigenvectors, $v_{\mathfrak n}(t,t_0)$, are aligned with those at
reference time slice, $t = t_{\rm ref}$, and they provide information as to
the optimal interpolating operator for the state ${\mathfrak n}$, namely $\sum_i
v^i_{\mathfrak n} {\cal O}^{\dagger}_i$.  
Figure~\ref{fig:Sigma_840_PrinCorrPlots} shows fits of principal correlators for 
the ground state and eight excited states in the $\Sigma$ spectrum 
for irrep $H_g$, where each state is identified as 
having spin-parity $J^P=\frac{3}{2}^+$. 
These states will be discussed later when we show how their spins
and flavors can be identified and when we show that they form part of a pattern of states
within the $\Sigma$ spectrum.

As in Ref.~\cite{Edwards:2011jj}, we make extensive use of the
operator ``overlap factors'', 
\mbox{$Z^\mathfrak{n}_i \equiv \langle
\mathfrak{n} | {\cal O}_i^\dag | 0 \rangle$} 
that occur in the spectral decomposition of the matrices of correlation functions, 
\beq
	C_{ij}(t) = \sum_{\mathfrak n} \frac{ Z^{\mathfrak{n}*}_i Z^{\mathfrak n}_j}{2m_{\mathfrak n}} e^{-m_{\mathfrak n}t} .
\eeq
Using the orthogonality condition for the eigenvectors $v^{\mathfrak{n}\dag} C(t_0)
v^\mathfrak{m} = \delta_{\mathfrak{n,m}}$,
the overlap factors are related to the eigenvectors 
through
\mbox{$Z^\mathfrak{n}_i = \sqrt{2 m_\mathfrak{n}} e^{m_\mathfrak{n}
  t_0/2}\, v^{\mathfrak{n}*}_j C_{ji}(t_0)$}. 

\begin{figure}
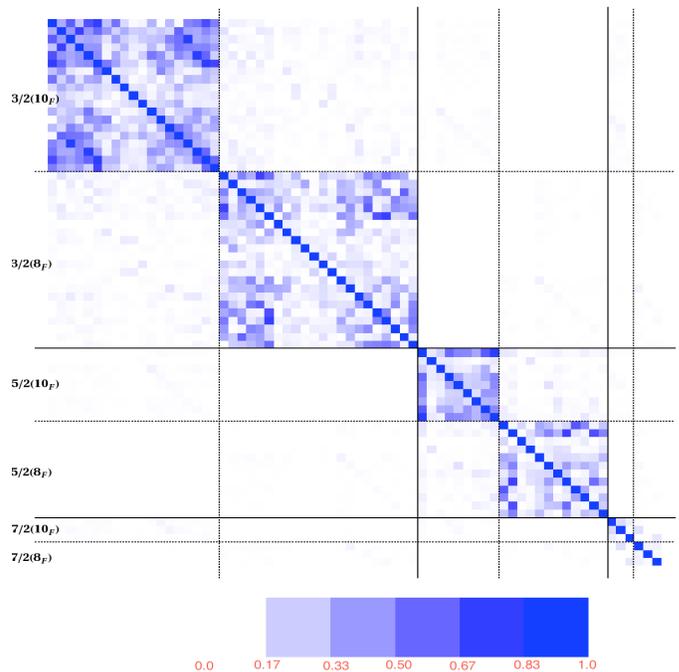

	     \ig[width=0.5\tw,height=0.5\tw]{Fig2_Corr_matrix_t5_plot.pdf} 
	\vspace{0.1in}
	\caption{Magnitudes of the elements of the correlator matrix,
	   $C_{ij}/\sqrt{C_{ii}C_{jj}}$, at time-slice 5 are shown for
	 $\Sigma$ at $m_{\pi} = $ 391 MeV according to the darkness scale at the 
	bottom. The blocks along the diagonal correspond to operators with the following spin-flavor
combinations: $\frac{3}{2}({\bf 10_F})$, $\frac{3}{2}({\bf 8_F})$, 
$\frac{5}{2}({\bf 10_F})$, $\frac{5}{2}({\bf 8_F})$, 
$\frac{7}{2}({\bf 10_F})$ and $\frac{7}{2}({\bf 8_F})$. 
	\label{fig:Sigma_840_t5_pixel_plot} 
	   }
	\end{figure}

        The baryon operators used in this work are constructed in two 
stages. The first stage is to construct operators in the continuum that have definite spin 
quantum numbers. In principle, these operators would produce a matrix of correlation 
functions that is orthogonal, i.e., proportional to $\delta_{J,J'}$, 
where $J$ and $J'$ are the spins of the source and sink operators.  In 
the second stage of the construction, the operators are subduced to the irreps 
of the cubic group so that they can be used on a finite lattice.  We observe 
that after the subduction, there remains a remarkable degree of rotational 
symmetry in the matrices of correlation functions that are obtained 
for baryons.  They exhibit
 approximately the orthogonality property that holds in the continuum. 
        An example of this is shown in 
	Fig.~\ref{fig:Sigma_840_t5_pixel_plot}, where the $\Sigma$ correlator 
	matrix, $C_{ij}(t)$, is shown at t=5 for irrep $H_g$, after normalizing each operator
	so that the diagonal elements are equal to one. 
	 The matrix indices range over 68 operators as follows: 
	19 $J=\frac{3}{2}$, ${\bf 10_F}$ operators;
	22  $J=\frac{3}{2}$, ${\bf 8_F}$ operators; 
	9 $J=\frac{5}{2}$, ${\bf 10_F}$ operators; 
	12 $J=\frac{5}{2}$, ${\bf 8_F}$ operators; 
	3 $J=\frac{7}{2}$, ${\bf 10_F}$ operator; and 
	3 $J=\frac{7}{2}$, ${\bf 8_F}$ operators.  
	The matrix is approximately block-diagonal with respect to $J$. 
Similar block-diagonal matrices of correlation functions are observed within other irreps and 
	they stem from the construction of operators starting from definite 
continuum spins.
	These approximately orthogonal matrices are key to the identification of the continuum spin of states.

\begin{figure*}
     \ig[width=1.0\tw,height=0.6\tw]{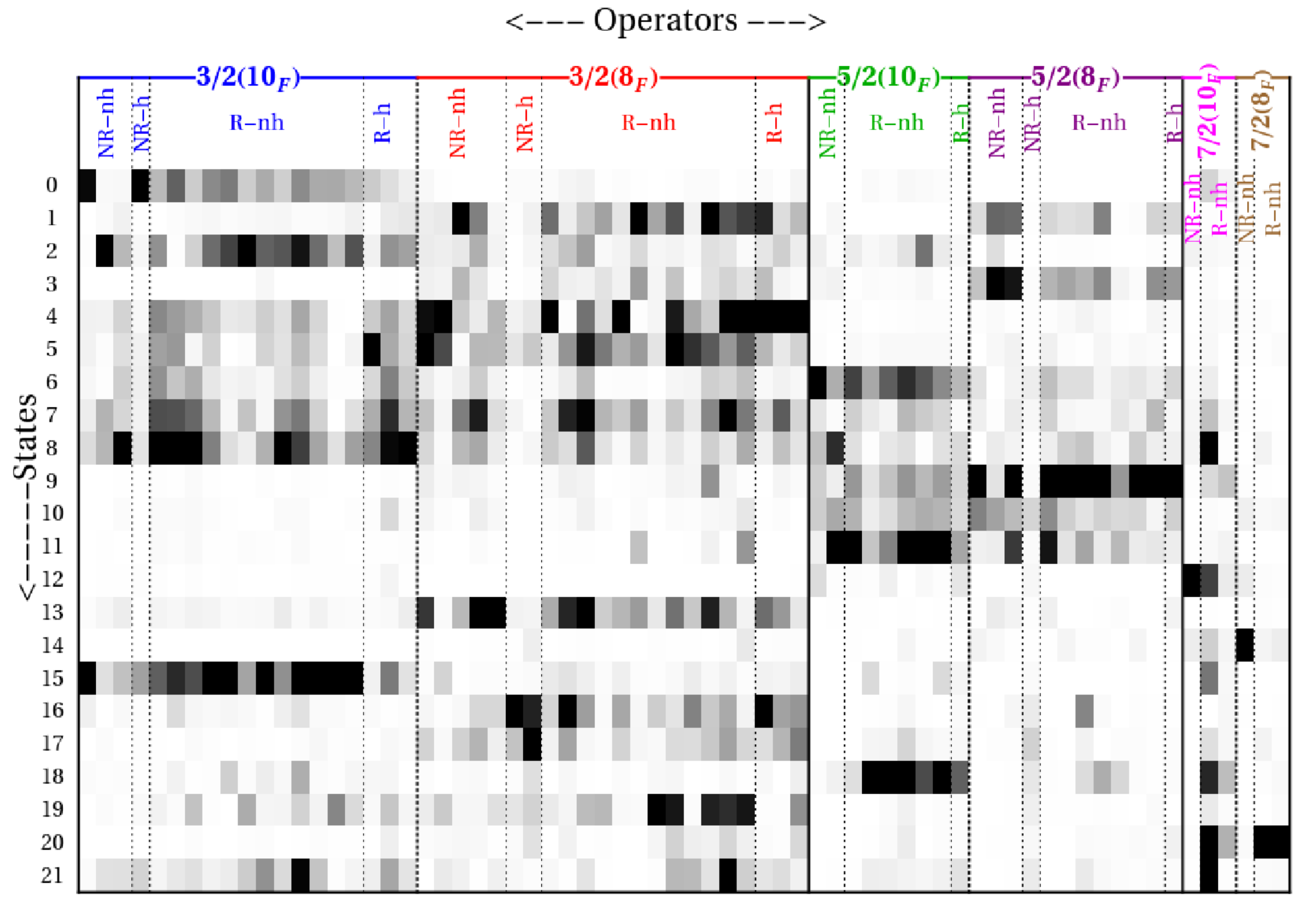} 
 \caption{
``Matrix" plot of values of operator overlaps, $Z^{\mathfrak{n}}_i$, for state ${\mathfrak{n}}$ and
operator $i$, normalized according to 
$\frac{Z^{\mathfrak{n}}_i }{ \mathrm{max}_\mathfrak{n}\left[ Z^{\mathfrak{n}}_i \right]}$ 
so that the largest overlap across all states for a given operator is unity.
For each of the $\mathfrak{n}$ = 0 to 21 states of $\Sigma$ in the $H_g$ irrep, 
the magnitude of each operator's overlap is shown 
for $m_{\pi} = $ 391 MeV.  Darker pixels indicate larger values of the operator overlaps
as in Fig.~\ref{fig:Sigma_840_t5_pixel_plot}.
Column labels indicate non-relativistic (NR) and relativistic (R) operators,
as well as hybrid (h) and non-hybrid (nh) operators.  In addition the
flavor irrep is indicated as (${\bf 10_F}$) for decuplet or (${\bf 8_F}$) for octet and continuum spins
of the operators are shown by 
$\frac{3}{2}$, $\frac{5}{2}$ and $\frac{7}{2}$. 
State 0, the ground state, and excited states 2, 8 and 15 are identified as $J^P= \frac{3}{2}^+$ states 
with dominant decuplet flavor symmetry.
States 1, 4, 5, 7 and 13 are identified as $J^P= \frac{3}{2}^+$, excited states 
with dominant octet symmetry. States 16 and 17 are $J^P= \frac{3}{2}^+$ excited states
with strong hybrid content and dominant octet flavor symmetry.
 \label{fig:Sigma_840_states_operators} 
   }
\end{figure*}

	Although flavor symmetry remains a broken symmetry in the continuum 
	because the strange quark in our calculations is heavier than the $u$ and $d$
	quarks, the correlator matrix is approximately block diagonal  
	with respect to flavor. As shown in Ref.~\cite{Dudek:2010wm}, which 
uses the same lattices, the breaking of $SU(3)_F$ symmetry is weak also 
for the mesons.
When the operators of a given flavor symmetry
	are dominant, we exploit that feature to identify the dominant flavor composition of a state,
	and when hybrid operators play a substantial role we use that to identify
	states with strong hybrid content.
	Figure~\ref{fig:Sigma_840_states_operators} illustrates how such identifications are made for states in the $H_g$ irrep of
	the $\Sigma$.   The plot shows the overlaps, $Z_i^{\mathfrak n}$, of operators 
	that create each state, with operators grouped by their continuum spin and flavor as indicated
	in the column labels at the top.  The columns also are labeled by whether the operators
	are based on relativistic (R) or non-relativistic (NR) Dirac spinors, or 
	hybrid constructions (h) or non-hybrid ones (nh).  
The operator
overlaps provide a reasonable identification of the spin, flavor and 
hybrid content of the $\Sigma$ states in irrep $H_g$ of the cubic group.  In particular,
the identifications of states with spin-parity $\frac{3}{2}^+$, and flavor irreps
${\bf 8_F}$ or ${\bf 10_F}$, are noted in the caption.
 Similar identifications are
made to extract the patterns of other states in our spectra.

 \section{Results} \label{sec:results}
In the following, we present the spectra in
 units of the $\Omega$ mass. 

\begin{figure*}
 \ig[width=\tw,height=1.4\tw] {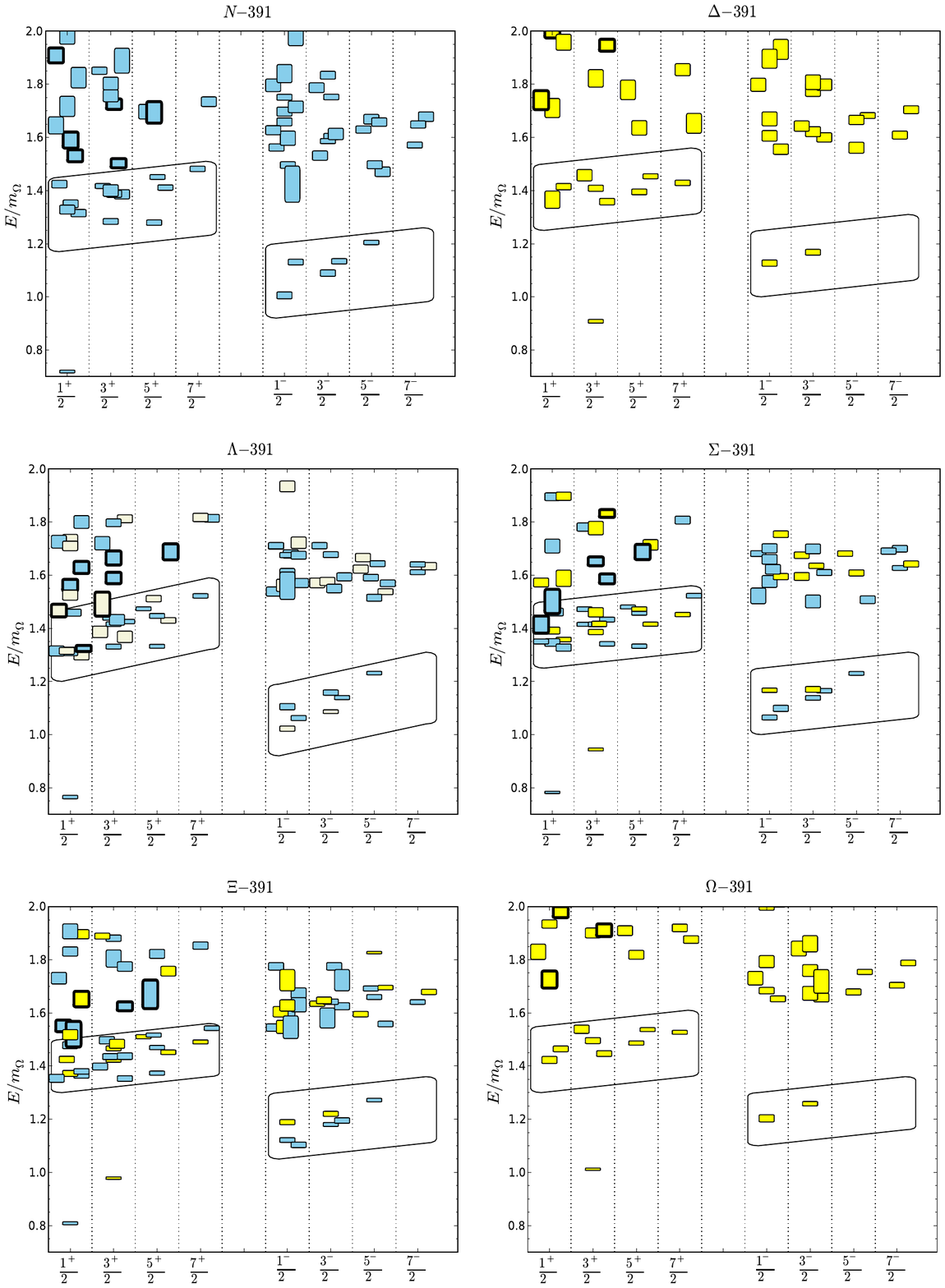}
\vspace{-1.2in}
\caption{ \label{fig:840_SU6xO3_spectra}  Results for baryon excited states using the ensemble with $m_{\pi}$ = 391 MeV are shown versus $J^P$.
Colors are used to display the flavor symmetry of dominant operators as follows: 
blue for ${\bf 8_F}$; beige for  ${\bf 1_F}$; yellow for ${\bf 10_F}$. 
Symbols with thick border lines indicate states with strong hybrid content.  
Calculations are for a $16^3\times 128$ lattice.  The lowest bands
of positive- and negative-parity states are highlighted within slanted boxes. 
The eight excited states of $\Sigma$, with $J^P = \frac{3}{2}^+$, that are shown 
within a slanted box, are $H_g$ states 1, 2, 4, 5, 7, 8, 13 and 15. Fits for
the same states are shown in 
Fig.~\ref{fig:Sigma_840_PrinCorrPlots} and identifications of their spins and flavors are 
noted in Fig.~\ref{fig:Sigma_840_states_operators}.  }
\end{figure*}

\begin{figure*}
 \ig[width=\tw,height=1.4\tw] {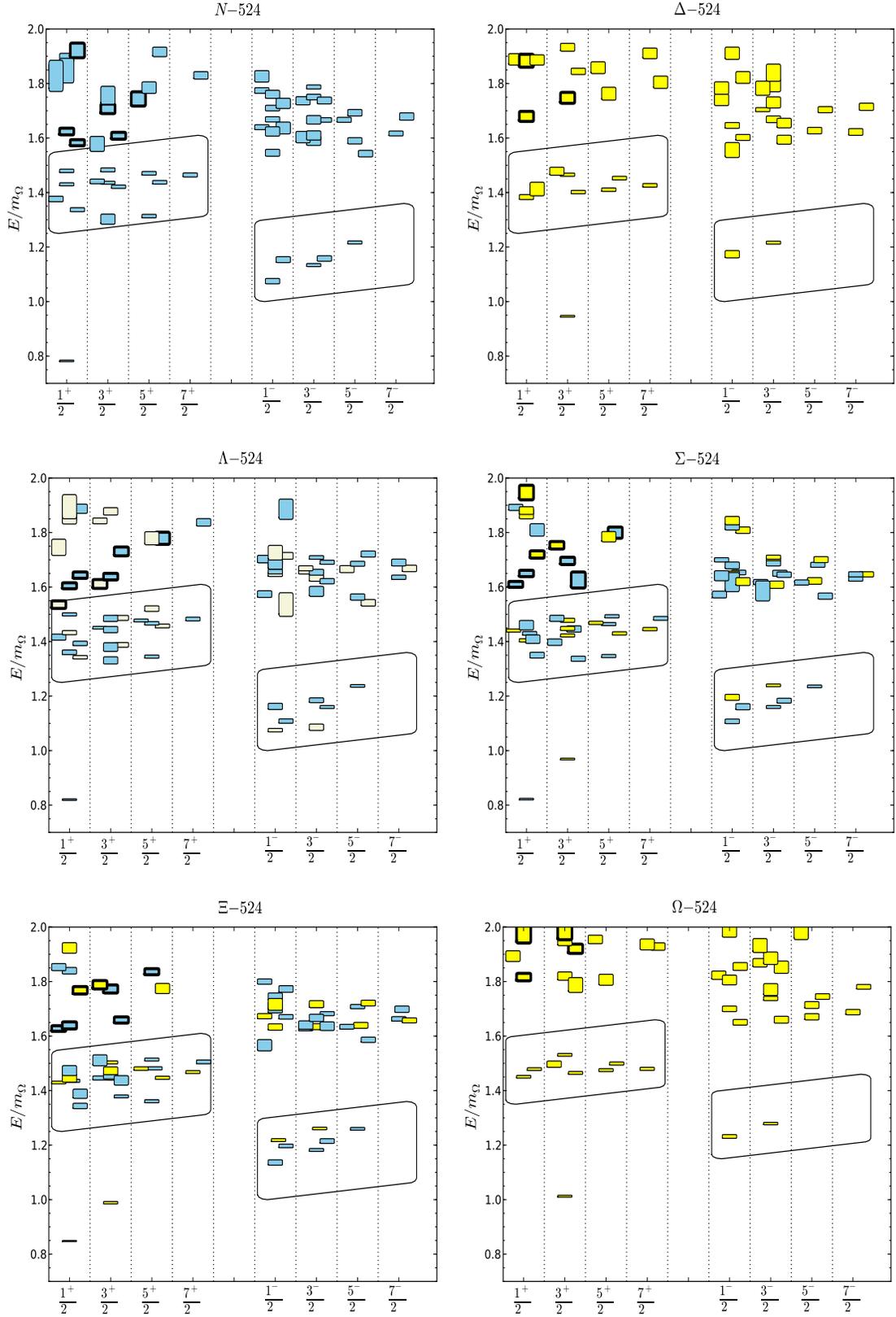}
\vspace{-1.2in}
 \caption{ \label{fig:808_SU6xO3_spectra}  Results for baryon excited states 
using the ensemble with $m_{\pi}$ = 524 MeV are shown versus $J^P$.
Symbols are as described in Fig.~\ref{fig:840_SU6xO3_spectra}. }
\end{figure*}

\begin{figure*}
%\vspace{-.6in}
 \ig[width=\tw,height=1.4\tw] {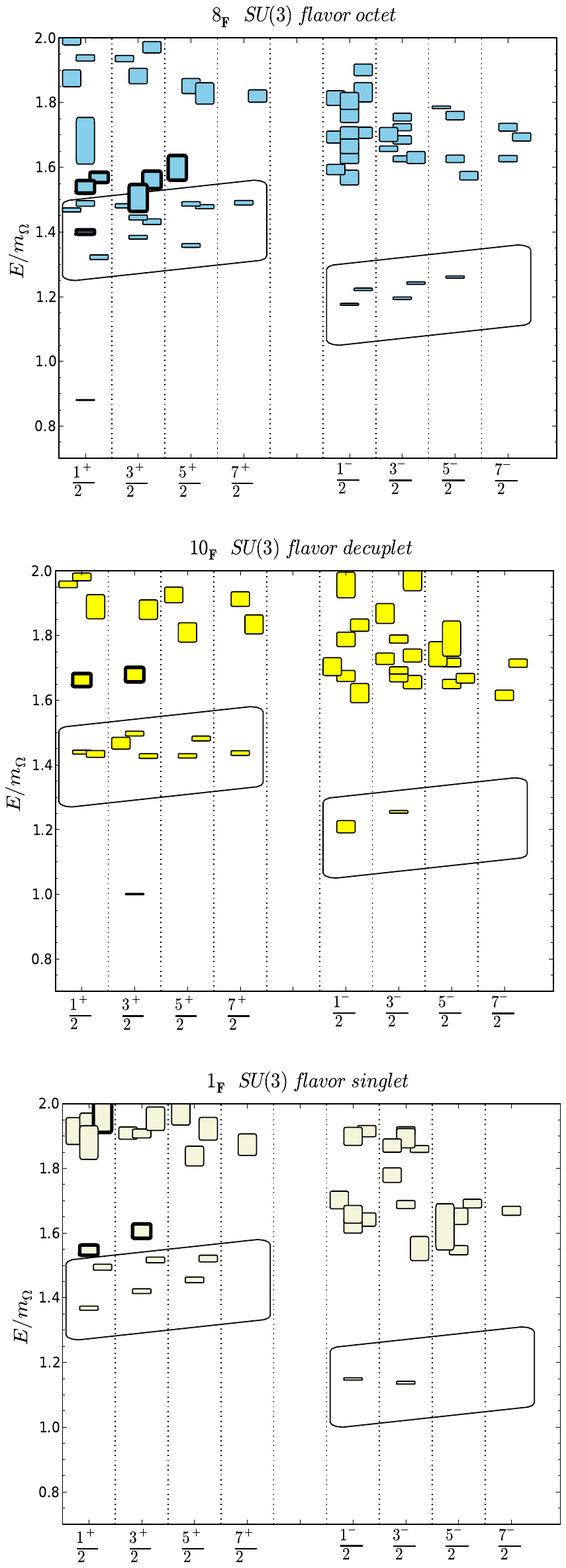}
\vspace{-1.2in}
 \caption{ \label{fig:743_SU6xO3_spectra}  Results for baryon excited states in flavor irreps
${\bf 8_F}$, ${\bf 10_F}$ and $\bf{1_F}$ obtained using the flavor-symmetric
point, with $m_{\pi}$ = 702 MeV, are shown versus $J^P$.
Symbols are as described in Fig.~\ref{fig:840_SU6xO3_spectra}. }
\end{figure*}

 The spectra of $N$, $\Delta$, $\Lambda$, $\Sigma$, $\Xi$ and
 $\Omega$ families of baryons for spins up to $\frac{7}{2}$ and both parities are shown at two 
different pion masses in Figures~\ref{fig:840_SU6xO3_spectra} and \ref{fig:808_SU6xO3_spectra}.  
The dominant 
flavor irrep is indicated by color: blue for ${\bf 8_F}$, 
yellow for ${\bf 10_F}$, and beige for ${\bf 1_F}$.
At the
$SU(3)_F$-symmetric point, the spectra are shown in Fig.~\ref{fig:743_SU6xO3_spectra},
now classified according to their $SU(3)_F$ flavor irrep: ${\bf 8_F}$, ${\bf 10_F}$ or ${\bf 1_F}$.
Results for the positive parity $N$ and $\Delta$ spectra 
 were presented earlier in ref.~\cite{Dudek:2012ag}. The $\Omega$ masses determined with the
operators used in this work differ by about 1\% from the values given in Table~\ref{tab:lattices}, which is within the
estimated uncertainty.
%%%%%%%%%%%%%%%%%%%%%%%
%%%%%%%%%%%%%%%%%%%%%%%

%%%%%%%%%%%%%%%%%%%%%%%
%%%%%%%%%%%%%%%%%%%%%%%

%%%%%%%%%%%%%%%%%%%%%%%
%%%%%%%%%%%%%%%%%%%%%%%

A general feature of
the lattice spectra is that there are bands of excited states with alternating parities for each family of baryons:  
the lowest band of states for each parity is shown inside the slanted boxes.  
In the lowest negative-parity band, the 
numbers of states of each spin and flavor agree with the expectations shown in Table~\ref{tab:D1_SU6_O3},
the table that shows the quantum numbers allowed by  
$SU(6)\times O(3)$ symmetry for one-derivative operators. 

In the lowest positive-parity band, the numbers of states for each spin and flavor agree with the
expectations shown in Table~\ref{tab:D2_SU6_O3}, the table that shows 
the quantum numbers allowed by $SU(6)\times O(3)$ symmetry for 
two-derivative operators. 
While at the $SU(3)_F$ symmetric point the counting involves
a single flavor irrep, for the $\Lambda$, $\Sigma$ and $\Xi$, the correct numbers
of states are obtained from summing the numbers for each of the two flavor irreps involved. 
This agreement between spectra and the expectations based on non-relativistic quark spins 
provides a clear signature of $SU(6)\times O(3)$ symmetry in the spectra. 

As noted in the caption of Fig.~\ref{fig:Sigma_840_states_operators}, the 
overlaps of the different operators classified according to their
flavor structure provides a means of identifying the dominant flavor
composition of the states in the spectra. 
Although the mixing of flavors increases as one moves away from the $SU(3)_F$-symmetric point, 
the dominant flavor identifications are shown by the
different colors used in the plots of spectra. It is clear from our analysis that 
for the mixed-flavor states, the $\Lambda$, $\Sigma$ and $\Xi$, we find spectra
that exhibit the multiplicities expected from exact $SU(3)_F$ flavor symmetry for each of
the flavor-identified multiplets. 
In summary, the baryon spectra are remarkably consistent with the $SU(6)\times O(3)$
expectations. It is this symmetry that is the basis for the quark model and the lowest
bands of lattice states include the same quantum numbers that occur in the 
quark model~\cite{Greenberg:1964,Isgur:1977ef,Isgur:1978xj}.

   In addition to the three-quark states that correspond to the quark model, 
a number of states are identified by their strong overlaps with hybrid interpolating 
operators, implying that they have a strong hybrid content.  These states, in which 
the gluons play a substantive role, are shown for positive parity by symbols with thick borders in 
Figs.~\ref{fig:840_SU6xO3_spectra}, \ref{fig:808_SU6xO3_spectra} and \ref{fig:743_SU6xO3_spectra}.
The two J = $\frac{3}{2}^+$, $\Sigma$ states in Fig.~\ref{fig:840_SU6xO3_spectra} with strong hybrid content and masses near 
 1.6$m_{\Omega}$ are $H_g$ states 16 and 17 in Fig.~\ref{fig:Sigma_840_states_operators}.
Note that the states with strong hybrid content generally are at high energy, typically 
about 0.7$m_{\Omega}$, or more, above the ground state.   

  In this work we have no three-derivative operators. 
Because of that, and because the relativistic operators generally play an important role at higher energies,
it is not meaningful to interpret the  
observed states at higher energies in terms of $SU(6)\otimes O(3)$ symmetry.
However, one sees qualitative similarity of the higher bands in Figs.~\ref{fig:840_SU6xO3_spectra} and
 \ref{fig:808_SU6xO3_spectra} to the corresponding bands found
at the flavor-symmetric point in Fig.~{\ref{fig:743_SU6xO3_spectra}.  

 It also is not meaningful to identify states with strong hybrid content 
in our negative-parity spectra and none are shown. Signals for states with significant hybrid 
content exist in negative-parity states at high energy, but they are based on the 
relativistic operators, whereas the three-derivative, non-relativistic operators that are absent  
may be equally, or more, important. Without a clear understanding of the relative importance of all the relevant
operators, we cannot identify negative-parity states with strong hybrid content.

The baryon spectrum does not admit the ``spin-parity exotics''
that provide a useful indication of hybrid states in the meson spectrum, and indeed we observe
that hybrid operators can have a significant, albeit not dominant, contribution
to many states.  Some examples can be found in Fig.~\ref{fig:Sigma_840_states_operators},
notably the lowest-lying spin-$\frac{3}{2}$ decuplet.  

The patterns and multiplicities of positive-parity states with strong hybrid content 
can be compared with the expectations based on non-relativistic quark
spins that are listed in Table~\ref{tab:D2_hybrid}.  
At the flavor symmetric point with $m_{\pi}$ = 702 MeV, which is shown in Fig.~\ref{fig:743_SU6xO3_spectra},
 and at $m_{\pi}$ = 524 MeV, which is shown in Fig.~\ref{fig:808_SU6xO3_spectra}, one sees all 
the positive-parity hybrid states corresponding to Table~\ref{tab:D2_hybrid}. At $m_{\pi}$ = 391 MeV,
which is shown in Fig.~\ref{fig:840_SU6xO3_spectra},
most of the hybrid states are seen. One decuplet, $J=\frac{1}{2}^+$ hybrid state is missing for $\Sigma$.
An octet, $J=\frac{3}{2}^+$ state and a decuplet, $J=\frac{3}{2}^+$ state are missing for $\Xi$.  These baryons involve 
mixings of ${\bf 8_F}$ and ${\bf 10_F}$ flavor symmetries and their spectra are particularly dense when subduced to the lattice irreps.
For example, an $H_g$ state must be found for each $J=\frac{3}{2}$, $\frac{5}{2}$ and $\frac{7}{2}$ state in 
the spectrum.  The $\Xi$, $J=\frac{5}{2}$ state with strong hybrid content
was found as the 29$^{th}$ state in the $H_g$ spectrum and was the highest state determined.  
The fact that a few of 
the states observed at higher values of the pion mass are not found at our lowest value of $m_{\pi}$ may be
because not all states have been determined. 
Overall, the lowest states with strong hybrid content are in reasonable accord with 
the expectations based on Table~\ref{tab:D2_hybrid}. 

  The excited states in the lowest bands of negative-parity are particularly well determined:
Fig.~\ref{fig:octet_u_840} shows the ones that are created predominantly 
by flavor-octet operators and Fig.~\ref{fig:decuplet_u_840} 
shows the ones that are created predominantly by flavor decuplet and singlet operators.
Note that we show the data for the mass of state $\mathfrak n$, namely $m_{\mathfrak n}$, 
in physical units that are obtained from the formula,
$m_{\mathfrak n} = 1672.45 \frac{m_{\mathfrak n, \rm latt}}{m_{\Omega, \rm latt}}$,
where $m_{\Omega, \rm     latt}$ is the $\Omega$ mass on the ensemble as given
< in Table~\ref{tab:lattices}.  Thus, 
the physical $\Omega$ mass is used to set the scale.
The patterns of these states are very similar for the different baryons.
For the flavor-octet states, there are two $\frac{1}{2}^-$ states, two $\frac{3}{2}^-$
states and one $\frac{5}{2}^-$ state. For each baryon, the energy
increases with spin $J$ and the highest energy is about 300 MeV above
the lowest energy, independent of the baryon.
For the flavor decuplet case, there is one $\frac{1}{2}^-$ state
and one $\frac{3}{2}^-$ state with 
about 70 to 100 MeV splitting, $\Sigma_{10}$ being an exception.  The flavor singlet
case has the same pattern except that the energies are lower and the
splitting is larger.

In a previous analysis of the nucleon spectrum using $N_f=2$ QCD, we obtained five low-lying 
negative-parity states in the lattice irreps $G_{1u}$, $H_u$ and $G_{2u}$~\cite{Bulava:2009jb}. 
 They could be 
interpreted as two $N(\frac{1}{2}^-)$ states, two $N(\frac{3}{2}^-)$ states and one 
$N(\frac{5}{2}^-)$ state, thus agreeing with the present work. A later analysis based on
$N_f = 2+1$ QCD obtained 
a sixth low-lying state, namely a third $N(\frac{3}{2}^-)$~\cite{Bulava:2010yg}, however 
that extra state was not as well determined.  Both of the mentioned works yielded two low-lying
$N(\frac{1}{2}^-)$ states, as does this work. The extra low-lying $N(\frac{3}{2}^-)$ 
state is not obtained in this work. We conclude that the third $N(\frac{3}{2}^-)$
state is spurious, and that the low-lying spectrum has a total of five negative-parity states, 
with strong evidence for low-lying bands consistent with SU(6)xO(3) symmetry.
 
Reference~\cite{Mahbub:2012ma} provides the masses of a few
low-lying, negative-parity states of the nucleon based on several pion masses, including
$m_{\pi}$ = 156 MeV.  
For $m_{\pi}$ values close to 400  MeV, there is good agreement with our 
masses for the two lowest $N(\frac{1}{2}^-)$ states shown in Fig.~\ref{fig:octet_u_840}. 
Reference~\cite{Menadue:2012mb} provides masses for several 
low-lying, negative-parity $\Lambda$ states. Lower pion masses were used in Ref.~\cite{Menadue:2012mb}, 
however, for the three lowest $\Lambda(\frac{1}{2}^-)$ states obtained at $m_{\pi} \approx 280$ 
MeV, there is acceptable agreement with our results at $m_{\pi} = 391$, namely for the  
$\Lambda_1(\frac{1}{2}^-)$ state in Fig.~\ref{fig:decuplet_u_840} and the two
$\Lambda_8(\frac{1}{2}^-)$ states in 
Fig.~\ref{fig:octet_u_840}.  Reference~\cite{Engel:2012xg} also provides masses for three 
low-lying $\Lambda(\frac{1}{2}^-)$ states at several values of $m_{\pi}$ using a larger 
lattice volume.  Those results are reasonably 
consistent with the masses of our three lowest-lying $\Lambda(\frac{1}{2}^-)$ states.

We note that there are several important limitations of the 
present study. They have been discussed in Ref.~\cite{Edwards:2011jj} and we conclude 
with a brief summary of them.  The 16$^3\times$128 lattice used is small, with spatial dimensions of
about 1.9 fm on a side.  The pion masses used
are significantly larger than the physical mass.  No operators that
efficiently couple onto scattering states (e.g., $\pi N$) are included. Studies of the resonances
that correspond to the three-quark states will require improvements that overcome
each of these limitations.

%%%%%%%%%%%%%%%%%%%%%%%
\begin{figure}
 \ig[width=0.5\tw,height=0.5\tw] {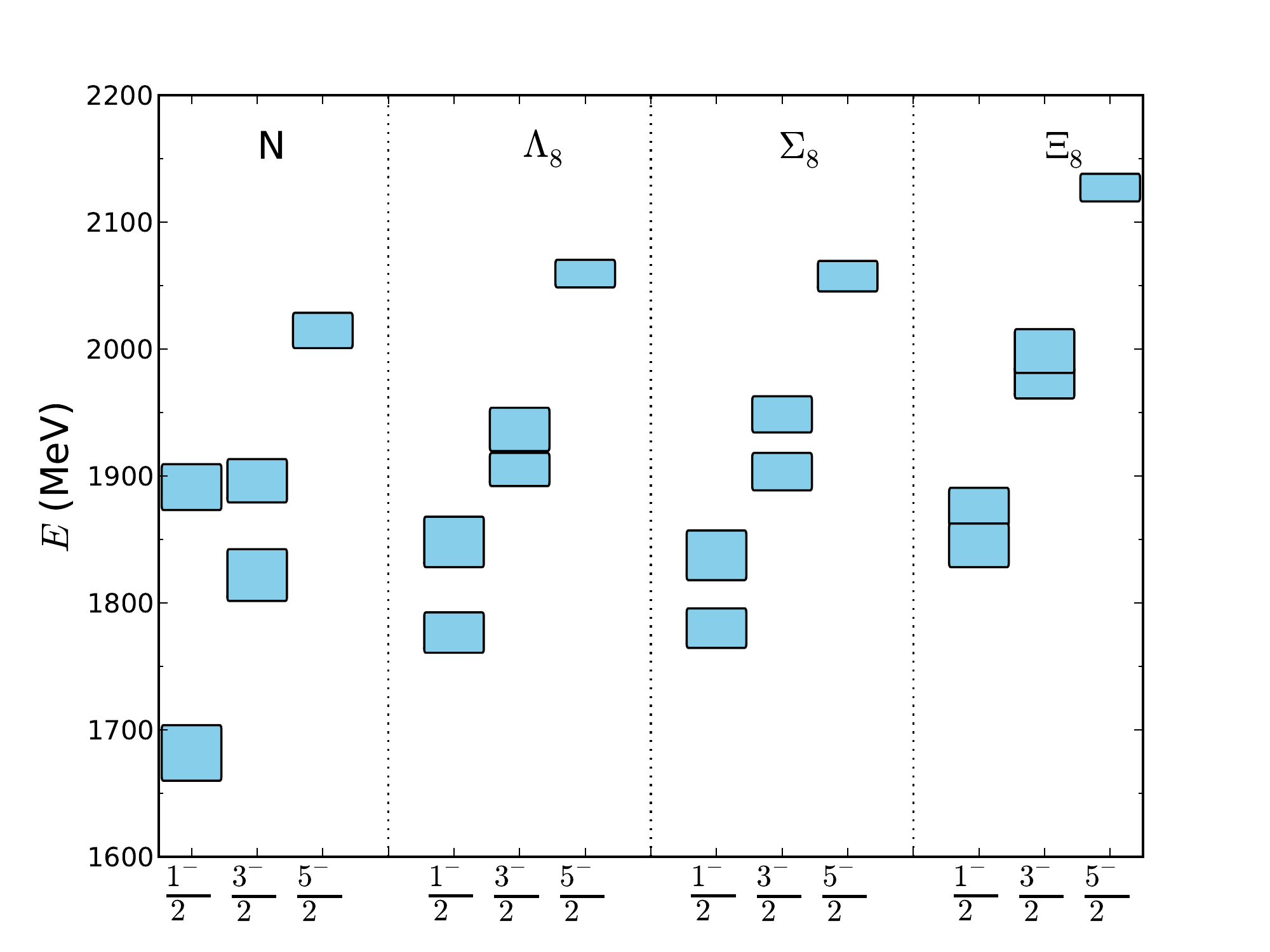} 
 \caption{ \label{fig:octet_u_840}  The lowest negative-parity states
 that are flavor-octet are shown for $m_{\pi}$ = 391 MeV.  }
\end{figure}
%%%%%%%%%%%%%%%%%%%%%%%

%%%%%%%%%%%%%%%%%%%%%%%
%%%%%%%%%%%%%%%%%%%%%%%
\begin{figure}
 \ig[width=0.5\tw,height=0.5\tw] {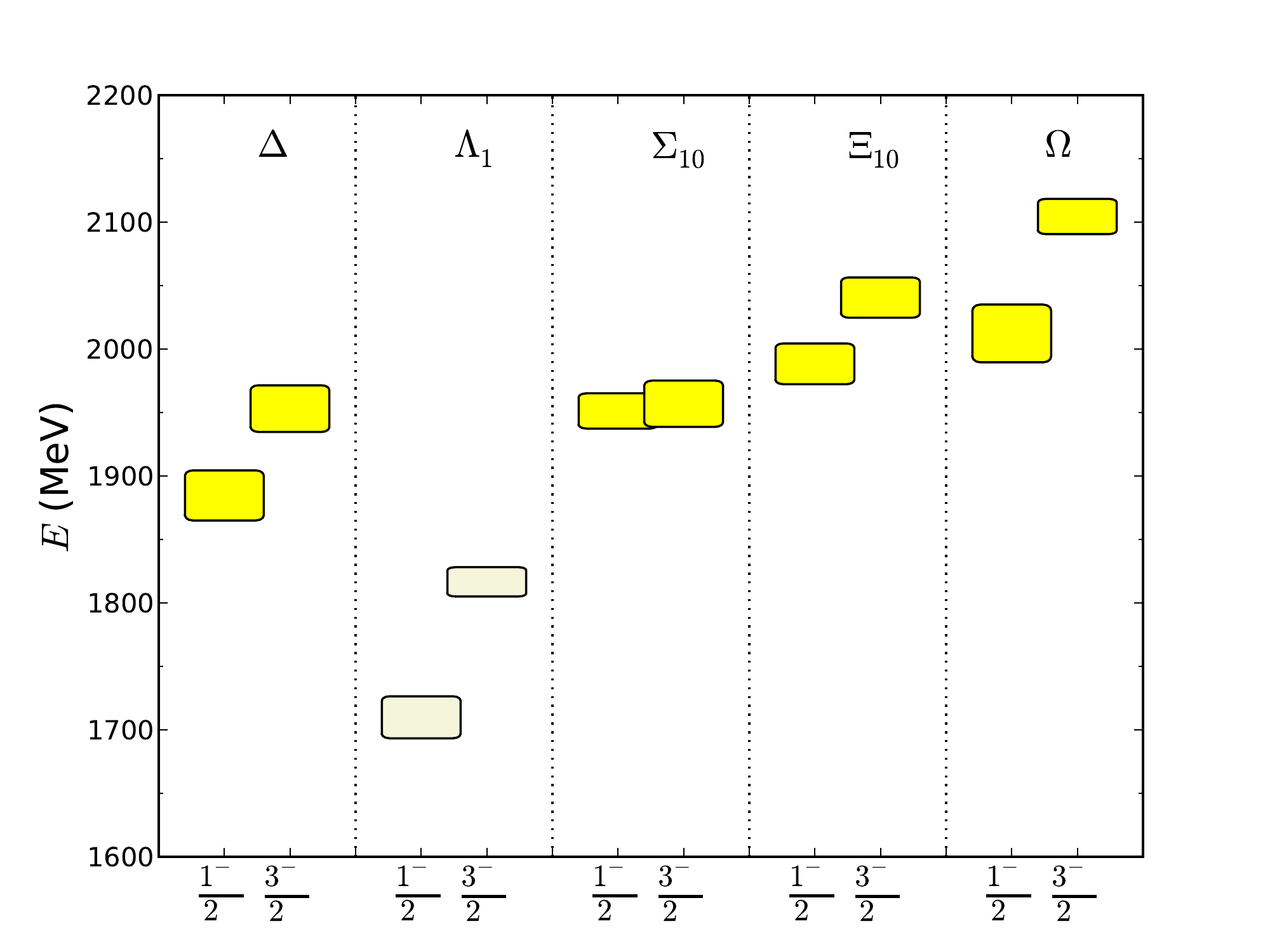} 
 \caption{ \label{fig:decuplet_u_840}  The lowest negative-parity states
that are flavor-singlet (beige) and decuplet (yellow) are shown for
$m_{\pi}$ = 391 MeV. }
\end{figure}
%%%%%%%%%%%%%%%%%%%%%%%

%\input{summary}

\section{Summary} \label{sec:summary}
This work presents results for baryons based on lattice QCD 
using the 16$^3\times$128 anisotropic lattices that were developed in Ref.~\cite{Lin:2008pr}.
 Excited state spectra are calculated for baryons that 
can be formed from $u$, $d$ and $s$ quarks, namely the $N$, $\Delta$, $\Lambda$, 
$\Sigma$, $\Xi$ and $\Omega$ families of baryons, for two pion masses, 391 MeV, 524 MeV, and 
at the $SU(3)_F$-symmetric point corresponding to a pion mass of 702 MeV.

The interpolating operators used incorporate covariant derivatives in 
combinations that correspond to angular- 
momentum quantum numbers $L$ = 0, 1 and 2. The angular momenta are combined with quark spins to build
operators that transform according to good total angular momentum, $J$, 
in the continuum. As noted in earlier works, approximate rotational symmetry 
is realized at the scale of hadrons, enabling us to identify reliably 
the spins in the spectrum up to $J = \frac{7}{2}$ from calculations at a single lattice spacing.  

The operators we have employed are classified according to the  
irreducible representations of SU(3)$_F$ flavor.  At the pion masses used, 
the SU(3)$_F$ symmetry is broken only weakly and states in the spectra can 
be identified as being created predominantly by operators of definite flavor 
symmetry ${\bf 8}_F$, ${\bf 10_F}$ or ${\bf 1_F}$.   

We find bands of states with alternating parities and increasing energies.  
Each state has a well-defined spin and generally a dominant flavor content can be identified. 
The number of non-hybrid states of each spin and flavor in the lowest-energy bands is in 
agreement with the expectations based on weakly broken $SU(6)\otimes O(3)$ 
symmetry. These states correspond to the quantum numbers of the quark model.

Chromo-magnetic operators are used to identify states that have strong 
hybrid content. Usually these states are at higher masses, about 
0.7$m_{\Omega}$, or more, above the lowest non-hybrid states.  
There is reasonable agreement between the
number of positive-parity states with strong hybrid content and the expectations 
of Table~\ref{tab:D2_hybrid} that are based on non-relativistic quark
spins, although a few of the expected states are not found at the
lowest pion mass.

With the inclusion into our basis of multi-hadron operators, which
couple efficiently onto multi-hadron scattering states, we expect
to find an increased number of levels in the spectrum.
As demonstrated in Ref.~\cite{Dudek:2012xn}, using the technique of
moving frames where the total momentum of the system is nonzero, the
increased number of levels allows for the extensive mapping of
the energy dependence of scattering amplitudes, and hence, the
determination of resonances.
The prospect of determining the properties of resonances  
provides a strong motivation for continued work on the spectra of baryons.
 
\acknowledgments

We thank our colleagues within the Hadron Spectrum Collaboration. Particular thanks to C. Shultz for his updates to our variational fitting code. {\tt Chroma}~\cite{Edwards:2004sx} and {\tt QUDA}~\cite{Clark:2009wm,Babich:2010mu} were used to perform this work on clusters at Jefferson Laboratory under the USQCD Initiative and the LQCD ARRA project. 
Gauge configurations were generated using resources awarded from the
U.S. Department of Energy INCITE program at Oak Ridge National Lab,
the NSF Teragrid at the Texas Advanced Computer Center and the
Pittsburgh Supercomputer Center, as well as at Jefferson Lab. 
SJW acknowledges support from U.S. Department of Energy contract DE-FG02-93ER-40762.  RGE,  
and DGR acknowledge support from U.S. Department of Energy contract
DE-AC05-06OR23177, under which Jefferson Science Associates, LLC,
manages and operates Jefferson Laboratory. NM acknowledges support from
Department of Science and Technology, India, under grant No. 
DST-SR/S2/RJN-19/2007.

%\bibliography{bibliography} 

\end{document}